\documentclass[amssymb,12pt,longbibliography,nofootinbib]{revtex4-1}
\newif\ifARXIV
\newif\ifUSEBBL

\ARXIVtrue

\USEBBLtrue

\def\journal{}
\def\link{}
\def\journal{Advanced Quantum Technologies 2300431 (2024)}
\def\link{https://doi.org/10.1002/qute.202300431}

\usepackage{amsmath,amsthm,appendix,bm,graphicx,hyperref,mathrsfs,setspace,slashed,xcolor}

\def\VOL{{\rm Vol}}
\newcommand{\BRA}[1] {\langle  #1 |}

\newcommand{\KET}[1] {| #1 \rangle}

\begin{document}
\rightline{\href{\link}{\tt \journal}}
\ifARXIV
\title{Protocol for nonlinear state discrimination in rotating condensate}
\author{\textsc{Michael R. Geller}}
\fi
\ifARXIV
\affiliation{\footnotesize{Department of Physics and Astronomy, University of Georgia, Athens, Georgia 30602, USA}}
\fi
\date{June 21, 2024}
 
\begin{abstract}
\ifARXIV
\centerline{}
\fi
\vskip 0.5in
\centerline{\bf Abstract}
\vskip 0.1in
\begin{spacing}{0.9}
Nonlinear mean field dynamics enables quantum information processing operations that are impossible in linear one-particle quantum mechanics. In this approach, a register of bosonic qubits (such as neutral atoms or polaritons)  is initialized into a symmetric product state $\KET{\psi }^{ \! \otimes n } $ through condensation, then subsequently controlled by varying the qubit-qubit interaction. We propose an experimental implementation of quantum state discrimination, an important  subroutine in quantum computation,
with a toroidal Bose-Einstein condensate. 
The condensed bosons here are atoms, each in 
the same superposition of angular momenta
 0 and $\hbar$, encoding a qubit.
A nice feature of the protocol is that only readout of individual quantized circulation states (not superpositions) is required.
\end{spacing}
\vskip 0.5in
\ifARXIV
\centerline{}
\fi
\end{abstract}
\maketitle

\section{Introduction}

A variety of atomtronic architectures have been proposed for quantum computing and quantum technology applications \cite{200804439,210708561}. 
Two main Bose-Einstein condensate (BEC) types have been considered for realizing qubits: multi-component condensates and multi-mode condensates. Multi-component and spinor condensate approaches \cite{CiracPRA98,TianPRA03,ByrnesWenPRA12,14103602,150303841,240310102}
encode a single qubit in two (or more) metastable atomic states, such as spin or hyperfine levels, with all atoms in the same translational mode (for example the motional ground state). 
Multi-mode approaches \cite{10055487,AmicoAghamalyanSciRep14,AghamalyanNguyenNJP16} encode a single qubit using two (or more) translational modes in a scalar condensate, such as a BEC in a double-well trapping potential. In the limit where there are a large number of condensed bosons in each well, the system becomes equivalent to two (or more) BECs, each with a well defined phase, connected by tunneling barriers that act as Josephson junctions \cite{10055487,AmicoAghamalyanSciRep14,AghamalyanNguyenNJP16}.
Arrays of such BECs can be produced in optical lattices and are described by the 
Bose-Hubbard model \cite{AmicoAghamalyanSciRep14,AghamalyanNguyenNJP16}.
Another multi-mode approach, which we adopt here, uses circulating states in a ring geometry 
\cite{RyuSamsonNat2020,KapalePRL05,RyuPRL07,RamanathanWrightPRL2011,EckelNat14,KimZhuPRL2018,231105523}  for the translational modes.

\begin{figure}
\includegraphics[width=10.0cm]{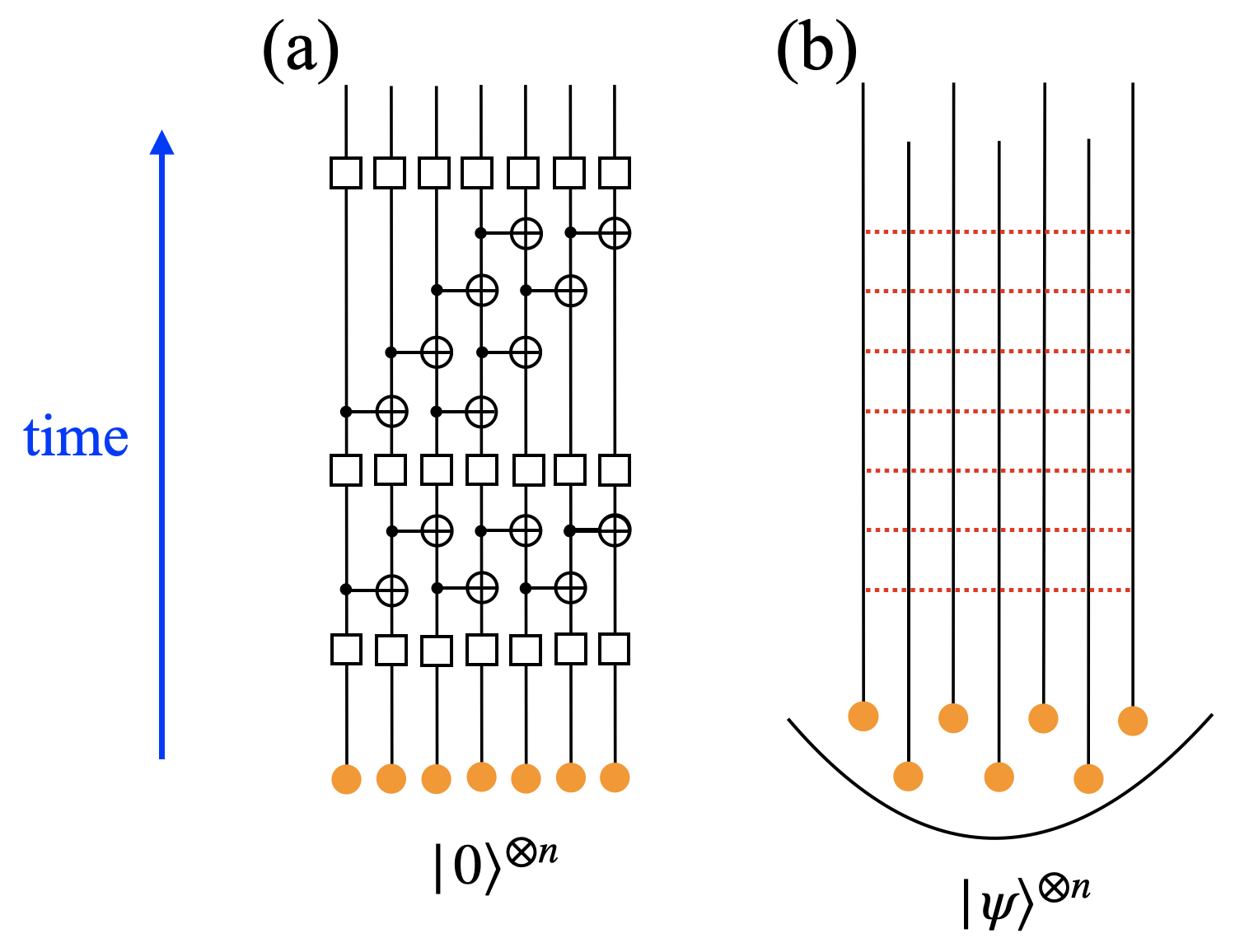} 
\caption{Nonlinear quantum information processing with a BEC. (a) In circuit-model quantum computation, a register of qubits is initialized to a product state (such as $\KET{0}^{ \! \otimes n} $), after which gates are applied, entangling the qubits. (b) In the nonlinear approach, the qubits ideally remain in a product state  $\KET{\psi(t)}^{ \! \otimes n} $ throughout the computation. The BEC simulates a single nonlinear qubit.}
\label{condensate computing figure}
\end{figure}

Given the demonstrated high performance and scalability of trapped ion qubits \cite{230503828}, superconducting qubits \cite{211203708,220706431,231105933}, and of neutral atom arrays \cite{BluvsteinNAT22,GrahamNat22}, what does a BEC qubit offer? 
We argue that it offers 
a platform for an alternative approach to quantum information processing that leverages the special properties of condensates. 
In this approach, a BEC is used to 
prepare a register of qubits in a product state $\KET{\psi}^{\! \otimes n} $ and control its subsequent evolution. From a quantum computing perspective, having multiple identical copies of an {\it unknown} input is already a useful resource (whereas classical information is freely cloned).
In standard circuit-model quantum computation, illustrated in Fig.~\ref{condensate computing figure}a, initialized qubits are subsequently entangled using two-qubit gates. 
Here we do the opposite and try to suppress entanglement, Fig.~\ref{condensate computing figure}b. This is achieved by making $n$ large, interactions weak, and by preserving permutation symmetry. In this limit entanglement monogamy \cite{CoffmanPRA00,220100366} bounds the pairwise concurrence to zero, and the BEC is exactly described by a nonlinear Schr\"odinger equation (the Gross–Pitaevskii equation \cite{GrossNuovoCimento1961,PitaevskiiJETP1961}), enabling novel dynamics \cite{PhysRevLett.81.3992,WuNiuPRA00,RiedelBohiNat2010,13030371,AmicoAghamalyanSciRep14,150706334,AghamalyanNguyenNJP16,RyuSamsonNat2020,XuPRR22,220613362,211105977}. 
The theory is developed in a large $n$ limit with a rigorous bound on the error resulting from the 
mean field approximation. {\it The nonlinear approach trades exponential time complexity for space complexity, requiring $n$ to be large.}

\clearpage

Does this mean that $n$ has to be exponentially large? Actually the requirements on $n$ are not that bad. This is because the BEC is assumed to be initialized in a product state, and it takes time 
$t_{\rm ent}$ for the atomic collisions to produce entanglement. Ideally, the whole experiment is performed in a short-time regime. We measure the accuracy of mean field theory by  $ \epsilon := \| \rho_{\rm eff}(t) - \rho_1(t) \|_1 $, and call this the model error. Here $\rho_{\rm eff}$  is the mean field state, $\rho_1$ is the exact state traced over all atoms but one, $t$ is the gate duration, and $ \| \cdot \|_1 $ is the trace norm. In a large family of condensate models  
\cite{ErdosJSP09,211209005}
\begin{eqnarray}
 \epsilon \le 
c  \frac{e^{t/ t_{\rm ent} }-1 }{n} ,
 \end{eqnarray}
where $c$  and $t_{\rm ent}$ are positive constants (model-dependent quantities independent of $t$ and $n$). Although the error might grow exponentially in time, there is always a short-time window $t < t_{\rm ent}$  where the required number of condensed atoms $ n \approx  c t / t_{\rm ent} \epsilon $ is sub-exponential in $t$.

We propose a demonstration of quantum information processing using this nonlinearity. In the remainder of this section we discuss the qubit encoding and state discrimination subroutine. The protocol is explained in Sec.~\ref{protocol section}. 
Conclusions are given in Sec.~\ref{conclusions section}, and additional information about the BEC  model and large $n$ limit are provided in the Appendix.

\clearpage

\subsection{Qubit encoding}

BEC-based qubits necessarily encode a small 
number of parameters ($ \psi_{0,1} \in {\mathbb C}$) into a large number of degrees of freedom and the map is not unique. However two encodings can often be considered:
Let $a_l^\dagger $ create an atom in BEC component $ l  \in \{0, 1\}$  (in a two-component condensate) or in translational mode $ l \in \{0, 1\}$ (in a two-mode condensate), and let $\psi_{0,1}$  be complex coordinates satisfying $| \psi_{0} |^2 + | \psi_{1} |^2  = 1$.  One encoding that is interesting from a  quantum foundations perspective is
\begin{eqnarray}
\KET{ {\rm CAT}_n } :=
\frac{ \psi_0 (a_0^\dagger)^n + \psi_1 (a_1^\dagger)^n }{\sqrt{n!}} \KET{ { \rm vac} } ,\ \ 
 \langle {\rm CAT}_n   | {\rm CAT}_n \rangle = 1,  \ \  
 n \ge 1,
 \label{cat encoding}
\end{eqnarray}
but this is a superposition of two macroscopically distinct BECs (a Schr\"odinger cat state) which would be highly susceptible to decoherence \cite{CiracPRA98}.
Instead we use the encoding 
\begin{eqnarray}
\KET{ F_n } :=
 \frac{( \psi_0 \, a_0^\dagger + \psi_1 \, a_1^\dagger)^n }{\sqrt{n!}} \KET{ {\rm vac }},  \ \ 
 \langle F_n  | F_n \rangle = 1,  \ \  
 n \ge 1,
\label{condensate encoding}
\end{eqnarray}
which is a condensate of $n$ bosons  $  \psi_0 \, a_0^\dagger + \psi_1 \, a_1^\dagger $, each a single atom in a superposition  of components or modes. (While $\KET{ {\rm CAT}_n } $ and  $\KET{ {F}_n } $ depend on both $n$ and $\psi_{0,1}$, the  latter dependence is suppressed.)
The encoding (\ref{condensate encoding}) was originally proposed by
Cirac {\it et al.}~\cite{CiracPRA98} and by
Byrnes {\it et al.}~\cite{ByrnesWenPRA12,14103602} for two-component condensates; in that case 
$\KET{ F_n } $ is a pseudospin coherent state \cite{14103602}.  But our
 $ a^\dagger_0 $
and $ a^\dagger_1 $ create atoms in circulating states of orbital angular momentum $0$ and 
$\hbar$, respectively, and it is better to regard  $\KET{ F_n } $ as a coherent 
state of atoms in angular momenta superpositions. 
The states (\ref{condensate encoding}) 
are mean field states since the atoms are not entangled. 
They  satisfy
\begin{eqnarray}
a_l \KET{ F_n }   = \psi_l \sqrt{n}   \KET{ F_{n-1} } \ \ \ {\rm and} \ \ \ 
a_l a_{l'}  \KET{ F_{n} } = \psi_{l} 
 \psi_{l'} \sqrt{n(n-1)}  \KET{ F_{n-2}} .
\label{condensate properties}
\end{eqnarray}
Because each atom in (\ref{condensate encoding})  
carries a copy of the qubit state $ \KET{\psi} = \psi_{0}  \KET{0} + \psi_{1} \KET{1} $, the state
$ \KET{ F_n } $ exhibits a bosonic orthogonality catastrophe \cite{200407166} in the large $n$ limit, meaning that close qubit states 
$ \KET{\psi} $
and
$ \KET{\psi^\prime}  $ encode to orthogonal 
$ \KET{ F_n } $  and $ \KET{ F^\prime_n } $ 
as $n \rightarrow \infty$ (the semiclassical limit in the spin coherent state picture \cite{14103602}).
Furthermore, due to the polynomial encoding in $ \KET{ F_n }$,
the single-particle superposition principle with respect to $ \psi_{0,1} $ is violated (see below).

\clearpage

In the atomtronic implementation we assume a toroidal BEC operated in a regime supporting metastable quantized circulation states with $l$ trapped vortices
\begin{eqnarray}
\KET{ \Phi_l^n} = \frac{(a_l^\dagger)^n }{\sqrt{n}} \KET{\rm vac} ,
\end{eqnarray}
where $a_l^\dagger$ creates an atom in the ring with angular momentum $l \in {\mathbb Z}$. 
These states are stabilized by the repulsive atomic interactions \cite{MuellerPRA02,BaharianPRA13}.
An atom with mass $m$ and $l=1$ has velocity  $ \hbar/ mR $ and circles the ring with angular velocity $ \Omega_0 = \hbar / mR^2 $. 
We construct a low-energy effective description for the BEC within the manifold of states (\ref{condensate encoding}). This is possible because they are selected out by the path integral in the large $n$ limit, due to their diverging contribution to the action. The action in the subspace spanned by these states is
\begin{eqnarray}
 S_{\rm eff} [ {\bar \psi}_l, \psi_l ]
= \int \! dt  \,
\BRA{ F_n }  i  \partial_t - H_{\rm rot}
\KET{F_{n}}   ,
\label{eff action definition}
\end{eqnarray}
where $ H_{\rm rot} $ is the BEC Hamiltonian in the rotating frame. The BEC is rotated with frequency  $ \Omega \approx  \Omega_0/2 $ to  bring the 0-vortex (no circulation) state $\KET{ \Phi_0^n} $ and the 1-vortex state $\KET{ \Phi_1^n} $ close in energy. Higher energy $l$ are then neglected, leading to a two-mode model. In the large $n$ limit (see Appendix) the saddle point equations are 
\begin{eqnarray}
 \frac{d}{dt} \! 
\begin{pmatrix} \psi_0 \\ \psi_1 \end{pmatrix}  
 \! = \!   - i H_{\rm eff} \!  \begin{pmatrix} \psi_0 \\ \psi_1 \end{pmatrix} \! , \ \ 
 H_{\rm eff}  = V_{01} \sigma^x + B_z \sigma^z + g (|\psi_0|^2 \! - \!  |\psi_1|^2)  \, \sigma^z  \!  .
 \label{mean field equations}
 \end{eqnarray}
The first two terms in $ H_{\rm eff} $ generate rigid $x$ and $z$ rotations of the Bloch sphere. Rotations about $x$ couple $l=0$ and $l=1$ angular momenta and are produced by breaking rotational symmetry.  Here $V_{01}$ is a matrix element for an applied potential energy barrier. 
The parameter $B_z$ is controlled by the frequency $\Omega$ of the BEC rotation discussed above. The nonlinear term describes a $z$ rotation with a rate that increases with increasing Bloch sphere coordinate ${\rm tr}(\rho \sigma^z) = |\psi_0|^2 \! - \!  |\psi_1|^2$, vanishes on the equator, and reverses direction for ${\rm tr}(\rho \sigma^z) < 0$. This  $z$-axis torsion \cite{MielnikJMP80}
(1-axis  twisting \cite{KitagawaUedaPRA1993}) of the Bloch sphere is the key to fast state discrimination, but is prohibited in ordinary single-particle quantum mechanics. Although the qubit here is informational and not associated with any physical 2-state system, we can define a logical basis $\{ \KET{0}, \KET{1} \} $ and treat it like any other qubit:
\begin{eqnarray}
\KET{\psi } = \psi_0  \KET{0} + \psi_1 \KET{1}
= \begin{pmatrix} \psi_0 \\ \psi_1 \end{pmatrix} \! , \ \ 
 \KET{0}  :=  \begin{pmatrix} 1\\ 0 \end{pmatrix} = \KET{ \Phi_{0}^n} , \ \ 
 \KET{1} :=  \begin{pmatrix} 0 \\ 1 \end{pmatrix} = \KET{ \Phi_{1}^n} .
 \label{qubit definition}
\end{eqnarray}
It should be emphasized that (\ref{condensate encoding}) is the physical state of the quantum gas,  not  (\ref{qubit definition}). 
However the basis states $ \KET{0} , \KET{1}  $ 
are the quantized circulation states $ \KET{ \Phi_{0,1}^n} $, which is important for the readout step.

\subsection{Single-input state discrimination}

As an application, we consider the problem of quantum state discrimination \cite{08101970,12042313,BaeJPA15,210812299}, a basic task in quantum information science. In the two-state variant considered here, a quantum state $ \KET{\psi} \in \{  \KET{a} ,  \KET{b}  \} $ is input to a processor, which knows the values of  $\KET{a}$ and $\KET{b}$ ahead of time and tries determine which was provided (with a bounded failure probability).
This is easy if $\KET{a}$ and $\KET{b}$ are orthogonal: For a qubit, a single unitary
$ U_{\rm read} = \KET{0} \BRA{a} + \KET{1} \BRA{b} $ rotates $ \alpha \KET{a} + \beta \KET{b} $ to $ \alpha \KET{0} + \beta \KET{1} $, which is then measured in the standard basis. The challenging case is when $\KET{a}$ and $\KET{b}$ are similar, $ | \langle a | b \rangle |^2 = 1 - 2^{-k}
 \  {\rm with} \  k \gg 1 $, where $n > 2^{k} $ identical copies of the input are required \cite{Helstrom1976}. 
 In minimum-error discrimination, the subroutine selects $\KET{a}$ or $\KET{b}$, each with some probability of error, and the 
objective is to minimize the average error. 
In unambiguous state discrimination, the subroutine  identifies $\KET{a}$ or $\KET{b}$  perfectly,  but has the possibility of abstaining, returning an 
inconclusive result.
State discrimination can be used to solve 
{\sf NP}-complete (and harder) problems \cite{PhysRevLett.81.3992,0502072,150706334}, at the expense of $2^{k}$ input copies and exponential runtime.  This cost reflects the limited information gained from measurement.  

Abrams and Lloyd \cite{PhysRevLett.81.3992} showed that certain nonlinearity in the Schr\"odinger equation would bypass this exponential cost, allowing {\sf NP}-complete problems to be solved efficiently (in an idealized setting with no errors or decoherence). But the presence of such nonlinearity would constitute a fundamental modification of quantum mechanics that is not supported by experiment \cite{BollingerPRL89,PhysRevLett.64.2261,WalsworthPRL90,MajumderPRL90}. 
In a condensate, the nonlinearity is not fundamental, but effective. Although we {\it can} realize nonlinear gates, this doesn't constitute a complexity violation, due to the large $n$ requirement of mean field theory. 

\clearpage

\section{Protocol}
\label{protocol section}

\begin{figure}
\includegraphics[width=10.0cm]{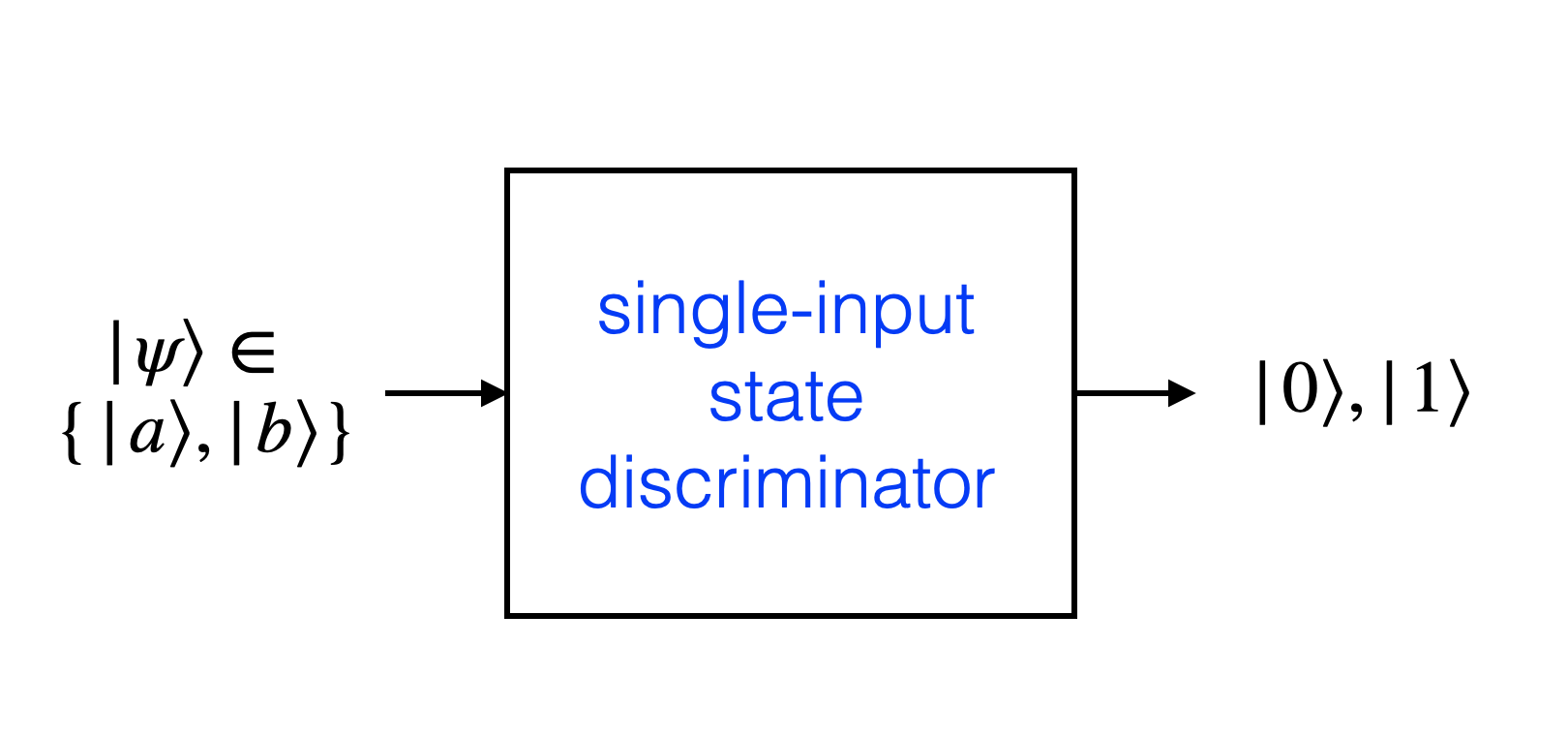} 
\caption{State discrimination channel. Here $\langle a | b \rangle  \neq 0 $ but  $\langle 0| 1 \rangle  = 0 $, so the channel must be nonlinear.
Note that the output is always a basis state, $\KET{0}$ or $\KET{1}$, simplifying readout.}
\label{discriminator figure}
\end{figure} 

The process is illustrated in Fig.~\ref{discriminator figure}. A single state  $ \KET{\psi} \in \{  \KET{a},  \KET{b} \}$ is input to the discriminator, which ideally returns output $ \KET{0}$ if  $ \KET{\psi} \! = \!  \KET{a}$, or returns $\KET{1} $ if  $ \KET{\psi} \! = \! \KET{b}$. 
The single-input discriminator regarded as a channel must be nonunitary, because the overlap 
$ \langle a | b \rangle  $ is not preserved.
Equivalently, the distance $ \| \rho_a - \rho_b \|_1$ between their density matrices in trace norm is not preserved in time (here $ \| X  \|_1 :=  {\rm tr} \sqrt{ X^\dagger X}   $). 
For pure states, $ \| \rho_a - \rho_b \|_1 = 2|\sin(\theta_{ab}/2) | $, where $\theta_{ab}$ is the angle between their Block vectors.
Linear completely-positive trace preserving (CPTP)  channels satisfy  $ \frac{d}{dt}  \| \rho_a - \rho_b \|_1 \le 0 $; they are either distance preserving or strictly contractive on the inputs \cite{RuskaiRMP1994}.
Because the discriminator orthogonalizes the potential inputs, it is {\it expansive} on those inputs:
 $ \frac{d}{dt}  \| \rho_a - \rho_b \|_1 > 0 $.
 Thus, the discriminator is described by a nonlinear PTP channel 
\cite{PhysRevLett.81.3992,MielnikJMP80,211105977}.

The implementation proposed here does not discriminate an unknown input (produced by a previous computation), but instead uses a black box state preparation step to randomly prepare  $ \KET{a} $ or $ \KET{b} $, with a small Bloch vector angle $ \theta_{ab} \ge 0$ between them. Then $ | \langle a | b \rangle |^2 = \cos^2(\theta_{ab} / 2) 
\approx 1 - (\theta_{ab} / 2)^2  $. 
This  can be accomplished by initializing in the $ \KET{\Phi^n_0} $ state and using $V_{01}$ and $B_z$ in (\ref{mean field equations}) 
to apply $x$ and $z$ rotations. 
(Ideally, this step is hidden from the remainder of the experiment.)  
The discrimination gate itself follows Refs.~\cite{PhysRevLett.81.3992,150706334}
and uses the $z$-axis torsion to increase the angle between $ \KET{a} $ and $ \KET{b} $. It's clear that  $ \KET{a} $ and $ \KET{b} $ should begin with equal and opposite $z$ components 
$z_a = - z_b$ [here $r^\mu_{a,b} = {\rm tr} ( \rho_{a,b}\sigma^\mu), \ \mu \in \{1,2,3\} $].
Consider a simple option with $ y_{a,b} = 0$, namely
\begin{eqnarray}
\KET{a}  &=&   \cos\bigg( \frac{\pi-\theta_{ab}}{4} \bigg)  \KET{0}   +      \sin\bigg(  \frac{\pi-\theta_{ab}}{4}  \bigg)  \KET{1} , \\
\KET{b}  &=&   \cos\bigg(  \frac{\pi+\theta_{ab}}{4}   \bigg)  \KET{0}  +   \sin\bigg(  \frac{\pi+\theta_{ab}}{4}  \bigg)  \KET{1} ,
\end{eqnarray}
which has 
\begin{eqnarray}
x_a = x_b =  \bigg| \! \cos \bigg( \frac{  \theta_{ab}} {2} \bigg)   \bigg| , \ \  
y_a = y_b = 0, \ \ 
z_{a} =  \sin \bigg( \frac{  \theta_{ab}} {2} \bigg), \ \ 
z_{b} =  -\sin \bigg( \frac{  \theta_{ab}} {2} \bigg).
\label{initial coordinates}
\end{eqnarray}
After switching on $g$, the two input options
evolve as $ R_z( \pm  gt \theta_{ab} ) $ and orthogonalize after a time $t  \approx \pi / g  \theta_{ab} $.  However this implementation does not have a favorable scaling with $ \theta_{ab} $.  
The optimal protocol for nonlinear discrimination was derived by Childs and Young (CY) in  \cite{150706334}.  Instead of (\ref{initial coordinates}), the CY gate 
begins with 
\begin{eqnarray}
 x_{a} = x_{b} 
= \bigg|  \! \cos\bigg( \frac{  \theta_{ab}} {2} \bigg) \bigg|, \ \ 
y_{a}= z_{a} = \frac{ \sin ( \frac{  \theta_{ab}} {2})}{\sqrt 2}, \ \  
y_{b}= z_{b} = - \frac{ \sin ( \frac{  \theta_{ab}} {2})}{\sqrt 2} ,
\end{eqnarray}
and applies $x$ rotations  to hold $ y_{a,b} = z_{a,b}$ during the subsequent evolution in order to reach antipodal points on the Bloch sphere. The options orthogonalize in a time $t = O( \log  \frac{1}{\theta_{ab} } )$, after which a readout gate
 $U_{\rm read}$ (defined with respect to 
 time-evolved $\KET{a}, \KET{b}$)  transforms them to circulation states $ \KET{ \Phi_{0}^n} $ or  $ \KET{ \Phi_{1}^n} $, which are then measured   via time-of-flight \cite{PhysRevLett.95.063201,MoulderBeattiePRA12}.
 
In an idealized context where (\ref{mean field equations}) is regarded as exact, and where there are no control errors, readout errors, decoherence errors, or noise, the nonlinear discriminator works perfectly every time. We refer to this idealization as a  {\it single-input} discriminator to distinguish it from the more familiar minimum error and unambiguous discriminators based on linear CPTP channels
\cite{08101970,12042313,BaeJPA15,210812299}.
Of course any actual atomtronic realization is likely to suffer from all such errors, and may fail to give the correct answer or return an answer at all. Although the theoretically achievable performance 
depends sensitively on the system and device details, and is beyond the scope of this work, we note that combining torsion with non-CP dissipation is predicted to implement an {{\it autonomous} discriminator \cite{211105977}, whose control sequence and operation is (mostly) independent of $\KET{a}$ and $\KET{b}$. In this implementation, the nonlinearity and dissipation 
create two basins of attraction with a shared boundary in the Bloch ball, one with an attracting fixed point near $\KET{0}$ and the other with an attracting fixed point near $\KET{1}$, 
giving the discriminator a degree of intrinsic fault-tolerance.

\begin{figure}
\includegraphics[width=14.0cm]{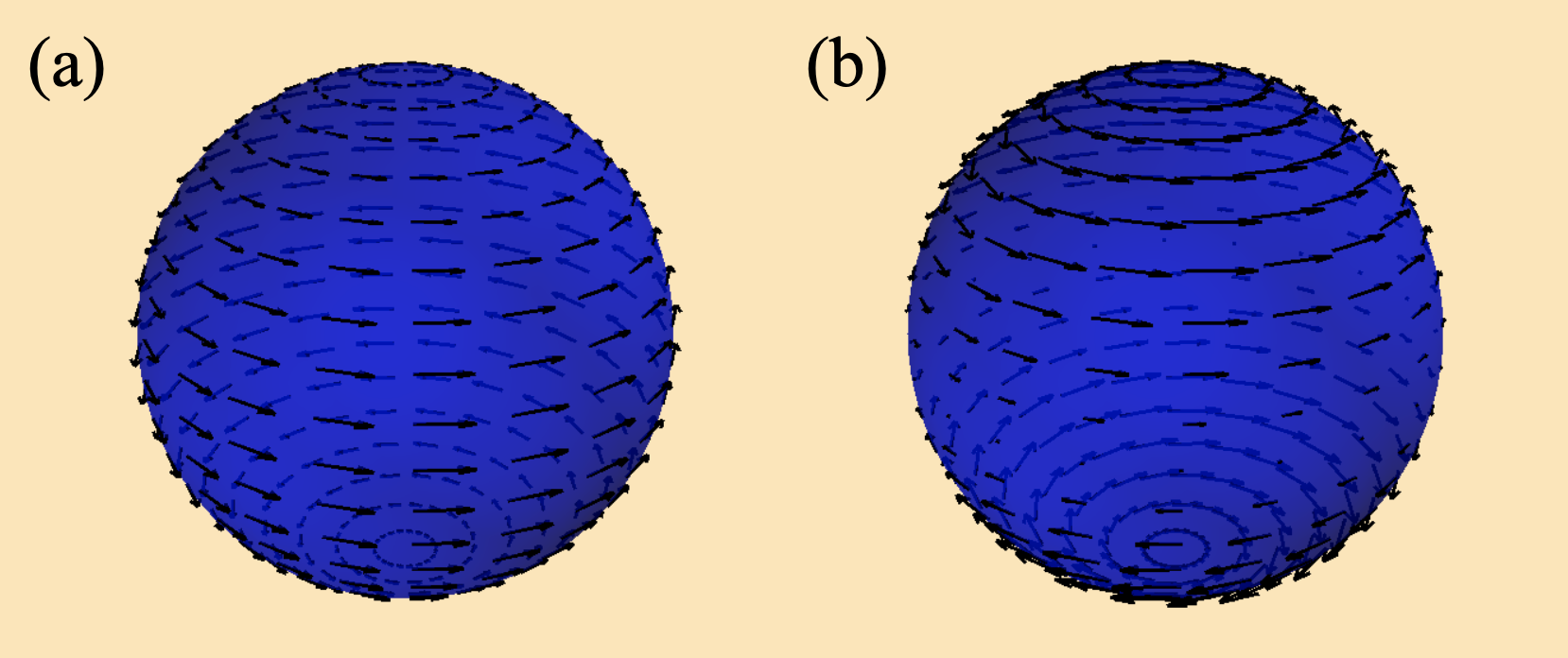} 
\caption{Linear versus nonlinear qubit evolution. (a) Unitary evolution. Vectors show the isometric flow of states on the Bloch sphere generated by linear Hamiltonian $H = \sigma^z$. 
(b) Torsion dynamics generated by nonlinear Hamiltonian 
$H = \BRA{\psi} \sigma^z \KET{\psi} \, \sigma^z$. }
\label{rotation vs torsion figure}
\end{figure} 

The presence of channel nonlinearity indicates a breakdown of the superposition principle. Figure~\ref{rotation vs torsion figure} illustrates a nice example of this effect:
In Fig.~\ref{rotation vs torsion figure}a, the evolution of a superposition 
$ \psi_0  \KET{0} +  \psi_1  \KET{1}  $
is given by a superposition of evolved basis states 
 $ e^{-it} \KET{0} $  and  $ e^{it} \KET{1} $,  shown as a velocity field. However in Fig.~\ref{rotation vs torsion figure}b, the evolved states $ e^{-it} \KET{0} $  and  $ e^{-it} \KET{1} $ are now static (phase factors are a global phase), whereas the actual dynamics is not,  except on the equatorial plane.

\section{Conclusions}
\label{conclusions section}
 
We have discussed an approach to quantum information processing that leverages the special properties of condensates, including their nonlinearity, and proposed an atomtronic implementation of a ``nonlinear'' qubit. An experimental demonstration of nonlinear state discrimination, while striking, would not by itself constitute a {\it computation}, because the qubit isn't coupled to anything. To implement a useful computation, the BEC qubit must be entangled with other qubits (for example trapped ions) in a scalable circuit-model quantum computer, which is not addressed here.

The standard models of quantum computation assume gates and errors based on linear CPTP channels.  Physical hardware, however,  might admit initial correlation and be better described by more general maps \cite{13120908,150305342}. 
It is therefore interesting to investigate any additional computational power enabled by quantum channels beyond the linear CPTP paradigm, as we did here. Another example 
was investigated by Chen {\it et al.}~\cite{220902481}, who experimentally demonstrated unambiguous state discrimination in a linear but non-Hermitian optical system. 
After completing this work, Gro{\ss}ardt posted a preprint \cite{240310102} proposing the use of a two-component BEC coupled to a neutral atom computer to simulate a large family of nonlinear Schr\"odinger equations.
Given their potential for fast quantum state discrimination and simulation of the nonlinear Schr\"odinger equation, the non-Hermitian and nonlinear approaches to quantum information processing deserve further exploration.

\acknowledgements

This work was partly supported by the NSF under grant no.~DGE-2152159.

\appendix

\section{BEC model}

Here we derive the qubit equation of motion (\ref{mean field equations}). We consider a toroidal BEC (thin circular ring with   radius $R$) with rotating tunneling barriers that act as Josephson junctions \cite{AmicoAghamalyanSciRep14,EckelNat14,AghamalyanNguyenNJP16,RyuSamsonNat2020}. Thin means the dynamics is quasi-1d in the azimuthal direction, $\theta$. This requires the energy, temperature, and effective interaction strength to be below an energy scale $\Delta \epsilon$ determined by the confining potential. The shape of the potential (without barriers) is mostly arbitrary as long as it is invariant under rotations about the axis threading the ring, which we call the $z$ axis.
The  angular momentum eigenfunctions on the ring are $\varphi_{l }(\theta) =  e^{ i l  \theta} / \sqrt{2 \pi R},$ with $l  \in {\mathbb Z}$ the angular momentum.  In the absence of the tunnel barriers and interaction, these are stationary states.
The condensate consists of $n$ weakly interacting bosonic atoms of mass $m$, each in their electronic ground state  $ \KET{\Psi_0}$.  At sufficiently low energy and densitiy, the atomic  collisions are elastic,  and the Hamiltonian is 
\begin{eqnarray}
H(t)   =   \int_{\VOL} \! \! d^3r  \,  \bigg\{ 
\frac{ \hbar^2 \nabla \phi^\dagger \!  \cdot  \! \nabla \phi }{2m} 
 + \frac{U}{2} \, \phi^\dagger  \phi^\dagger \phi \, \phi  
  + V  \phi^\dagger \phi
\bigg\}, \ \ [\phi({\bf r}) , \phi^\dagger({\bf r}') ] = \delta({\bf r}-{\bf r}').
\label{3d ring. hamiltonian}
 \end{eqnarray}
Here $\VOL$ is the volume of the ring, 
 $U  = 4 \pi \hbar^2 a_{\rm s} /m$ is a short-range interaction strength (proportional to the s-wave scattering length $a_{\rm s}$), and $V({\bf r},t) $ is a confining potential, including  the rotating barriers. Acting on the vacuum, $\phi^\dagger({\bf r})$ creates a bosonic atom in state $ \KET{\Psi_0}$ at point ${\bf r}$. We assume a tunable repulsive interaction with $a_{\rm s} \ge 0$. We also assume zero temperature, no dissipation, and no disorder.

Two rotating tunnel barriers  are used to implement an atomtronic quantum interference 
device \cite{AghamalyanNguyenNJP16,RyuSamsonNat2020}.
When the barriers are turned on, the Hamiltonian  (\ref{3d ring. hamiltonian}) is time dependent. Assuming the barriers are rigidly rotated about the $z$ axis with frequency $\Omega$, 
we have
$ V({\bf r},t) = e^{-i \Omega t L_z / \hbar}  \, V({\bf r}, 0) 
\, e^{i \Omega t L_z / \hbar} $,
where $L_z$ is the angular momentum.

However, we can transform to a noninertial reference frame in which the Hamiltonian, $H_{\rm rot}$, is time independent. 
Decomposing the time-evolution operator in the lab frame as 
\begin{eqnarray}
U_{\rm lab} = T e^{- \frac{i}{\hbar} \int_0^t  H dt'}  = e^{-i \Omega t L_z / \hbar} U_{\rm rot}  ,
 \end{eqnarray}
 we obtain
 \begin{eqnarray}
 \frac{dU_{\rm rot} }{dt} = - \frac{i}{\hbar}  H_{\rm rot}   U_{\rm rot} , \ \  H_{\rm rot} = e^{i \Omega t L_z / \hbar} ( H - 
 \Omega L_z ) e^{-i \Omega t L_z / \hbar}  = H(0) 
 - \Omega  L_z .
 \end{eqnarray}
 
Next we discuss the two-mode limit:
In the low energy, thin ring limit,  we can expand the field operators and angular momentum as
 \begin{eqnarray}
\phi({\bf r}) = \sum_l    \frac{e^{ i l  \theta} }{ \sqrt{ \VOL }}   \, a_l , \ \ 
L_z = \sum_l \hbar l   a^\dagger_l a_l , \ \ 
[ a_l  , a_{l '}^\dagger ] = \delta_{l  l '} ,
\end{eqnarray}
which leads to
\begin{eqnarray}
H_{\rm rot}   &=&    \frac{ \hbar \Omega_0}{2}  \sum_{l }  l^2 
a_l ^\dagger a_l 
+ \frac{U}{2 \, \VOL}
\sum_{l_1,  l_2, l_3}
 a_{l_1 + l_3 }^\dagger a_{l_2 -l_3}^\dagger a_{l_2 } a_{l_1} 
 +  \sum_{l_1,  l_2} \, V_{l_1   l_2}  \, a_{l_1}^\dagger a_{l_2}
- \hbar \Omega \sum_l l  a^\dagger_l a_l , \ \ \ 
\label{angular momentum hamiltonian}
\end{eqnarray}
where
\begin{eqnarray}
V_{l_1   l_2} &=&   \oint \! \frac{ d \theta}{2 \pi}  \, V(\theta, t \! = \! 0) \,  e^{-i (l_1 - l_2) \theta} .
\label{v matrix element}
\end{eqnarray}
Nonzero $ V_{l_1 l_2} $ induce transitions between angular momentum states. Then we have
\begin{eqnarray}
H_{\rm rot}   =    \sum_{l } \hbar \omega_l
a_l ^\dagger a_l 
+ \frac{U}{2 \, \VOL}
 \sum_{l_1,  l_2, l_3}
 a_{l_1 + l_3 }^\dagger a_{l_2 -l_3}^\dagger a_{l_2 } a_{l_1} 
 +   \sum_{l_1,  l_2} \, V_{l_1   l_2}  \, a_{l_1}^\dagger a_{l_2}
  - \frac{ n \hbar \Omega^2}{2  \Omega_0} ,
 \label{angular momentum hamiltonian rewritten}
\end{eqnarray}
where
\begin{eqnarray}
 \omega_l = \frac{(\Omega - l \Omega_0)^2 }{2 \Omega_0} , \ \ 
 \omega_0 = \frac{\Omega^2 }{2 \Omega_0} , \ \  \omega_1 = \frac{(\Omega - \Omega_0)^2 }{2 \Omega_0}.
\end{eqnarray}

As explained above, the BEC is rotated with frequency  $ \Omega \approx  \Omega_0/2 $ to  bring the $l=0$ and $l=1$ states close in energy. 
We restrict (\ref{angular momentum hamiltonian rewritten}) to angular momenta $l = 0,1$ neglecting the others on the basis of their higher energy. 
Then
\begin{eqnarray}
 \sum_{l_1,  l_2, l_3}
 a_{l_1 + l_3 }^\dagger a_{l_2 -l_3}^\dagger a_{l_2 } a_{l_1} 
& =& \sum_{l \in {\mathbb Z} } \bigg\{  
 a_l^\dagger a_{-l}^\dagger a_0 a_0
 +  a_{l+1}^\dagger a_{-l}^\dagger a_0 a_1
 +  a_{l}^\dagger a_{1-l}^\dagger a_1 a_0
 +  a_{1+l}^\dagger a_{1-l}^\dagger a_1 a_1
 \bigg\} \ \ \ \  \\
 & =& 
 a_{0}^\dagger a_{0}^\dagger a_0 a_0 
 +  a_{1}^\dagger a_{1}^\dagger a_1 a_1 
 + 4  a_{0}^\dagger a_{1}^\dagger a_1 a_0 \\
 & =& 
( a_{0}^\dagger a_0 )^2 - a_{0}^\dagger a_0 
+ ( a_{1}^\dagger a_1 )^2 - a_{1}^\dagger a_1 
+ 4  a_{0}^\dagger a_0 a_{1}^\dagger a_1 .
\end{eqnarray}
This leads to a two-mode model
\begin{eqnarray}
H_{\rm rot}  \! = \!   \sum_{l=0,1} \bigg( \! \hbar  \omega_l + V_{ll} + \gamma a_l ^\dagger a_l - \gamma \! \bigg) a_l ^\dagger a_l 
+ \gamma^\prime  a_{0}^\dagger a_0 a_{1}^\dagger a_1  + \bigg( \! V_{01} a_{0}^\dagger a_{1} \! + \! {\bar V_{01}}  a_{1}^\dagger a_{0}  \! \bigg)  - \frac{ n \hbar \Omega^2}{2  \Omega_0} , \ \ \ \ \ \ \ 
  \label{two-mode hamiltonian}
\end{eqnarray}
where 
\begin{eqnarray}
 \gamma = \frac{U}{2 \, \VOL}, \ \  \gamma^\prime =  4 \gamma = \frac{2 U}{\VOL} .
\end{eqnarray}
In what follows we will treat $\gamma , \gamma^\prime \ge 0$ as independent parameters, 
allowing 
(\ref{two-mode hamiltonian}) to apply to other systems as well.
The last term in (\ref{two-mode hamiltonian}) subtracts the classical kinetic energy of the spinning ring: $ n \hbar \Omega^2 \! / 2  \Omega_0 = \frac{1}{2} I_{\rm ring} \Omega^2, \ I_{\rm ring} = n m R^2 \! .$

Finally we discuss the large $n$ limit:  Condensates feature an enhanced two-particle interaction $ \langle a^\dagger a^\dagger  a a \rangle \approx n(n-1)$ caused by the effectively infinite-ranged interaction between condensed atoms. This makes a naive large $n$ limit unphysical, because the energy per particle diverges \cite{LiebSeiringerPRA2000}, and our low-energy assumptions would be violated.
The framework discussed here is instead based on  a modified large $n$ limit where the interaction simultaneously weakens as $1/n$ (a standard assumption in rigorous studies of mean field theory \cite{LiebSeiringerPRA2000,Benedikter2016}).
This allows for a rigorous study of the large  $n$ limit including bounds on the accuracy of mean field theory. For real $ V_{01} \! = \! \oint \! \frac{ d \theta}{2 \pi}  \, V(\theta) \,  e^{i \theta} $,  and after dropping the classical kinetic energy term (which does not affect the qubit dynamics) we obtain
 \begin{eqnarray}
H_{\rm rot}   =    \sum_{l=0,1} \bigg(  \! ( \! \hbar  \omega_l + V_{ll} ) a_l ^\dagger a_l  
+ \gamma a_l ^\dagger a_l^\dagger a_l a_l 
- \gamma  a_l^\dagger a_l \! \bigg)
+ \gamma^\prime  a_{0}^\dagger a_0 a_{1}^\dagger a_1  + V_{01}  \bigg( \! a_{0}^\dagger a_{1} \! + \! a_{1}^\dagger a_{0} \!  \bigg) . \ \ \ 
\label{hrot real v01}
\end{eqnarray}
Evaluating  (\ref{eff action definition}) and assuming $n \gg 1$ leads to (setting $\hbar = 1$)
\begin{eqnarray}
 S_{\rm eff} \! = \! n \!  \!  \int \!  \! dt 
\bigg\lbrace \! 
 \sum_{l=0,1} \! \bigg( \!  {\bar \psi}_l   i \partial_t  \psi_l
\! - \!  ( \! \omega_l + V_{ll} ) | \psi_l|^2
\! - \!   n  \gamma  \,  |\psi_l |^4 \! \bigg)
\! - \!  n  \gamma^\prime | \psi_0  \psi_1 |^2 
\! - \!   V_{01}  ( {\bar \psi}_0 \psi_1
\! + \!  {\bar \psi}_1 \psi_0) \! \bigg\rbrace . \ \ \ \ \ \ \ \ 
\label{raw effective action}
\end{eqnarray}
Here $ {\bar z}$ denotes complex conjugation.
Due to the $O(n^2)$ interaction energies in (\ref{hrot real v01}), we cannot take the $n \rightarrow \infty $ limit in (\ref{raw effective action}) without violating our low-energy assumptions. Instead we consider a modified limit defined by the simultaneous limits $ \gamma, \gamma^\prime \rightarrow 0 $, $ n \rightarrow \infty$, and low energy. To evaluate (\ref{raw effective action}) in this limit we assume that the interaction strengths decrease with $n$ as $\gamma = K / n$ and $\gamma^\prime = K^\prime / n $, where $K$ and $K^\prime$ are now fixed coupling constants. 
Then
\begin{eqnarray}
 S_{\rm eff} \! = \!  n  \!  \!  \int \!  \! dt 
\bigg\lbrace \! 
 \sum_{l=0,1} \! \bigg( \!  {\bar \psi}_l   i \partial_t  \psi_l
 \! - \!  ( \! \hbar  \omega_l + V_{ll} ) | \psi_l |^2
\! - \!   K    |\psi_l |^4 \bigg)
\! - \!  K^\prime | \psi_0  \psi_1 |^2 
\! - \!   V_{01}  ( {\bar \psi}_0 \psi_1
\! + \!  {\bar \psi}_1 \psi_0) \! \bigg\rbrace . \ \ \  \ \ \ \ \ 
\label{effective action}
\end{eqnarray}

To obtain (\ref{effective action}) we have used the results summarized below in Table \ref{state comparison table},  which also gives the corresponding results for encoding (\ref{cat encoding}).
In the large $n$ limit the stationary phase approximation leads to  (\ref{mean field equations}) with
\begin{eqnarray}
 H_{\rm eff} &  = & \begin{pmatrix}  \hbar \omega_0 +V_{00} +  2 K  |\psi_0|^2  + K^\prime \,  |\psi_1|^2    & V_{01}  \\  V_{01}   &  \hbar \omega_1  + V_{11}  +  2 K  |\psi_1|^2    +   K^\prime \,  |\psi_0|^2   \end{pmatrix} \\
 &=& V_{01} \sigma^x + B_z \sigma^z + g  \,  {\rm tr}(\rho \sigma^z) \,\sigma^z   + {\rm const.} , 
 \end{eqnarray}
where
\begin{eqnarray}
B_z :=  \frac{\hbar \omega_0 - \hbar \omega_1 + V_{00} - V_{11}}{2}  \ \ {\rm and} \ \
 g :=  \frac{2K - K^\prime}{2} .
\end{eqnarray}
This concludes the derivation of (\ref{mean field equations}).

\begin{table}[htb]
\centering
\caption{One- and two-particle correlators  versus encoding.}
\begin{tabular}{|c|c|c|}
\hline
    & $ \langle a_{l}^\dagger a_{l'} \rangle $  & 
    $ \langle a_{l}^\dagger a_l a_{l'}^\dagger a_{l'} \rangle $   \\
 \hline
 $ \KET{ {\rm CAT}_n } $  & 
 $n \,   |  \psi_l |^2 \delta_{l l'}$  & $n(n-1) \,   |  \psi_l |^2
\delta_{l l'}$ \\
\hline
 $ \KET{ {F}_n } $  & 
 $n \,    \psi_l^* \psi_{l'}$  & 
 $n(n-1) \,   |  \psi_l |^2
 |  \psi_{l'} |^2 $ \\
 \hline
\end{tabular}
\label{state comparison table}
\end{table}


\ifUSEBBL
\bibliography{Paper.bbl}
\else
\bibliography{/Users/mgeller/Dropbox/bibliographies/CM,/Users/mgeller/Dropbox/bibliographies/MATH,/Users/mgeller/Dropbox/bibliographies/QFT,/Users/mgeller/Dropbox/bibliographies/QI,/Users/mgeller/Dropbox/bibliographies/group,/Users/mgeller/Dropbox/bibliographies/books}
\fi

\end{document}